\documentclass[letterpaper,12pt]{article}
\usepackage{geometry}
\usepackage{fancyhdr}
\usepackage{authblk}
\usepackage{graphicx}
\usepackage{mathtools}
\usepackage{cite}
\usepackage{hyperref}
\usepackage{soul} 
\usepackage{color}
\usepackage{caption}
\usepackage{subcaption}

\begin{document}

\title{Random Raman lasing}

\author[1,2]{Brett H. Hokr}
\author[1]{Michael Cone}
\author[1]{John D. Mason}
\author[3]{Hope T. Beier}
\author[3]{Benjamin A. Rockwell}
\author[3]{Robert J. Thomas}
\author[4]{Gary D. Noojin}
\author[2]{Georgi I. Petrov}
\author[5]{Leonid A. Golovan}
\author[1,2]{Vladislav V. Yakovlev}

\affil[1]{\small{Department of Physics \& Astronomy, Texas A\&M University, College Station, TX, USA}}
\affil[2]{\small{Department of Biomedical Engineering, Texas A\&M University, College Station, TX, USA}}
\affil[3]{\small{711th Human Performance Wing, Optical Radiation Bioeffects Branch, JBSA Fort Sam Houston, TX, USA}}
\affil[4]{\small{TASC Inc., San Antonio, TX, USA}}
\affil[5]{\small{Faculty of Physics, M. V. Lomonosov Moscow State University, Moscow, Russia}}

\date{\small{\today}}

\maketitle


\textbf{Propagation of light in a highly scattering medium is among the most fascinating optical effect that everyone experiences on an everyday basis and possesses a number of fundamental problems which have yet to be solved~\cite{Mie1908,Pasternack1995,Licht1996,Tegen1996,Skipetrov2006,Yakovlev2010,Bertolotti2012}. Conventional wisdom suggests that non-linear effects do not play a significant role because the diffusive nature of scattering acts to spread the intensity, dramatically weakening these effects. We demonstrate the first experimental evidence of lasing on a Raman transition in a bulk three-dimensional random media. From a practical standpoint, Raman transitions allow for spectroscopic analysis of the chemical makeup of the sample~\cite{Freudiger2008,Schopf2002}. A random Raman laser could serve as a bright Raman source allowing for remote, chemically specific, identification of powders and aerosols. Fundamentally, the first demonstration of this new light source opens up an entire new field of study into non-linear light propagation in turbid media~\cite{Wiersma1997}, with the most notable application related to non-invasive biomedical imaging~\cite{Siddique1995,Song2010,Redding2012}.}

Raman scattering is the inelastic scattering of a photon off a molecule. The frequency of the scattered photon is determined by the frequency of a vibrational level of the molecule, making Raman spectroscopy a valuable tool for molecular and structural identification. In the context of Raman lasing, spontaneous Raman scattering and stimulated Raman scattering (SRS) are analogous to fluorescence and stimulated emission, respectively. However, the nonlinearity associated with SRS distinguishes it from stimulated emission~\cite{Shen1965}. In other words, the gain in the Raman scattered field is proportional to the intensity of the pump. Hence, Raman lasers typically require high optical pump intensities in order to attain sufficient gain to achieve lasing~\cite{Spillane2002,Rong2005b,Rong2005,Troccoli2005}. Unlike a traditional laser, whose wavelength is determined by the electronic structure of the gain medium, the wavelength of a Raman laser is determined not only by the vibrational spectrum of the gain medium, but also by the wavelength of the pump laser used.

Random lasing was first predicted by V. S. Letokhov in 1968~\cite{Letokhov1968}, but was not experimentally observed until 1994~\cite{Genack1994,Lawandy1994}. Random lasers operate on many of the same principles as traditional lasers. The exception is that in random lasers multiple elastic scattering provides feedback in place of mirrors. Analogous to traditional lasers, random lasers have a threshold where gain from electronic transitions characteristic of the material exceeds losses and exponential amplification occurs at the lasing frequency~\cite{Wiersma2008}. Random lasers can be divided into two categories, coherent and incoherent, based on the feedback that drives them~\cite{Cao2001}. Coherent feedback occurs with the generation of unstable periodic trajectories by multiple elastic scattering events~\cite{Cao2000}. These trajectories are analogous to many randomly oriented ring cavities existing in the gain medium. When multiple scattering acts only to return energy back into the gain medium, but phase information is lost in the process, the system is driven by incoherent feedback~\cite{Cao1999}. Incoherent feedback is diffusive in nature, meaning the wave properties of light are negligible in such a system. Thus, such a system can be accurately described with Monte Carlo simulations.

A random Raman laser has a primary gain mechanism of SRS, and relies on elastic scattering to provide feedback into the gain medium. To date, random Raman lasing has only been observed in a one-dimensional fiber system~\cite{Turitsyn2010}. While many properties of the random Raman laser closely mimic those of a traditional random laser there are some notable differences. First, the gain bandwidth of a Raman transition is quite narrow. This makes it very unlikely for the resonant frequency of a closed cavity to lie within this small gain window and observe gain. Thus, the dynamics of a random Raman laser will be dominated by incoherent feedback. Second, Raman transitions are much faster than electronic transitions, thus the transient (picosecond) dynamics of pulse propagation through the medium will play a pivotal role. The complicated dynamics of nonlinear pulse propagation in a turbid medium make an analytical approach to describing this problem very difficult. In order to better understand these processes, we employed a Monte Carlo model developed previously~\cite{Hokr2013b}. These simulations were used to guide our experiments and offer the chance to illuminate aspects of the dynamics that cannot be easily observed experimentally.

The random Raman laser (illustrated in Fig.~1a) was made of barium sulfate ($\mathrm{BaSO_4}$) powder with particle diameters of $1 - 5 \; \mathrm{\mu m}$. $\mathrm{BaSO_4}$ plays the role of both the Raman gain medium and the scattering centers, and was chosen because of its low absorption and high scattering cross section throughout the visible spectrum. Micron-sized particles were chosen to ensure that the Raman gain in any one particle is not too large, thus requiring the feedback provided via elastic scattering for sufficient amplification of the Raman signal. To pump the random Raman laser, laser radiation with a wavelength of $532 \; \mathrm{nm}$ and a pulse duration of $40 \; \mathrm{ps}$ was employed. The incident radiation was gently focused onto the sample using a slightly offset telescope as a compound lens, allowing for adjustments in the beam diameter incident on the sample (see Fig.~1b). The generated Raman output light was collected at near normal incidence and passed into an energy meter, a CCD camera, or a spectrometer via a lens for the appropriate measurements.

At low pump pulse energies, spontaneous Raman scattering dominates our detected Raman signal. However, once the pump energy increases beyond the threshold, ($1.05 \; \mathrm{mJ}$ in our experiment) gain exceeds losses and SRS dominates the conversion process, which is the onset of random Raman lasing (see Fig.~2a). At a maximum of $11.5 \; \mathrm{mJ}$ of pump energy, as much as $2.0 \; \mathrm{\mu J}$ of Raman signal was measured by the energy meter. Assuming a homogeneous angular distribution of emitted Raman photons, our conversion efficiency of pump photons into Raman photons was approximately $10^{-2}$. This demonstrates that the random Raman laser is remarkably efficient considering typical conversion efficiencies for spontaneous Raman scattering are on the order of $10^{-11}$~\cite{Boyd2003}.

Above the lasing threshold, the Raman spectrum narrows to a single Raman peak with a frequency shift of $985 \; \mathrm{cm}^{-1}$ as shown in Fig~2b. At higher pump energies, a second peak was observed corresponding to a second order SRS signal. This signal was generated by a second $985 \; \mathrm{cm}^{-1}$ shifted Raman response. The primary Raman light served as the pump for a second Raman process which corresponded to a total Raman shift of $1970 \; \mathrm{cm}^{-1}$. The presence of a second-order process further illustrates the substantial efficiency of this random Raman laser.

Below the lasing threshold, the spatial distribution of the emitted Raman light on the random Raman laser is very broad and predominately due to spontaneous Raman scattering (see Fig.~3). This is the result of deep penetrating photons that spend a long time in the medium, providing a greater chance to be converted into a Raman photon~\cite{Hokr2013b}. These deep penetrating photons have a larger probability of exiting the sample with large radial offsets. However, above the threshold, the majority of the energy is emitted from a highly localized area on the surface (see Fig.~3). This is primarily due to the fact that a large amount of the SRS light is generated near the surface, resulting in much smaller radial offsets. Strong SRS generation near the surface can be attributed to the slower speed of light diffusion in turbid media compared to the speed of light in a vacuum~\cite{Albada1991}. Thus, photons initially arriving in the medium do not have time to leave the front surface of the sample before photons from the back of the pulse enter, causing the pulse to effectively be compressed. This light compression raises the intensity, generating a large SRS signal close to the surface (See the supplimental material for a video of the simulation showing this effect).

We have demonstrated the first random Raman laser in a three-dimensional random medium. Above the threshold, SRS gain exceeds losses due to diffusion and absorption, resulting in Raman lasing. While many of the features are similar to traditional random lasers, the dynamics of random Raman lasing have a few notable differences which we have elucidated with the help of Monte Carlo simulations. These devices open up an entire new frontier on which to test the understanding of nonlinear optics, and offer a possible mechanism that could provide a luminously bright source of Raman light making remote sensing of powders or aerosols a natural extension.

\bibliographystyle{naturemag}
\bibliography{../../../references}

\textbf{Supplementary Information} is available in the online version of the paper.

\textbf{Acknowledgments} The authors acknowledge the Texas A\&M Supercomputing Facility (http://sc.tamu.edu/) for providing computing resources useful in conducting the research reported in this paper. This work was partially supported by National Institutes of Health Grant R21EB011703 and National Science Foundation Grants ECCS-1250360, DBI-1250361, and CBET-125036.

\textbf{Author Contributions} BHH, MC, JDM, HTB, BAR, GN, GIP, and VVY conducted experiments. BHH and RJT conducted simulations. BHH, BAR, GIP, LAG, and VVY designed research. All authors contributed to the writing of the paper.

\textbf{Author Information} Reprints and permissions information is available at \\www.nature.com/reprints. The authors declare no competing financial interests. Readers are welcome to comment on the online version of the paper. Correspondence and requests for materials should be addressed to VVY (yakovlev@bme.tamu.edu).

\begin{figure}[p]
\centering
\includegraphics[width=1.0\textwidth]{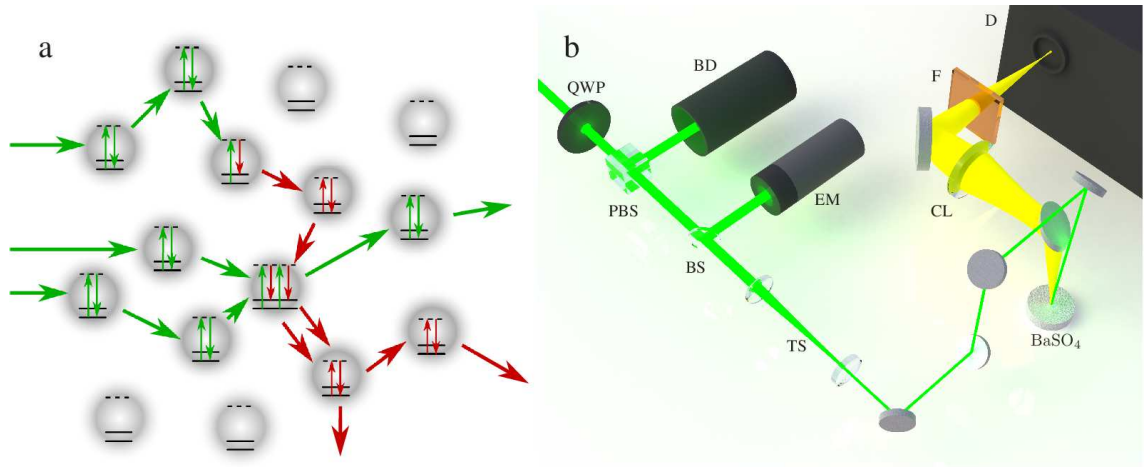}
\caption{\textbf{Schematics of the random Raman laser and experimental setup.} \textbf{a}, Conceptual drawing illustrating random Raman lasing that is built up from spontaneous Raman scattering. \textbf{b}, Simplified diagram of the experimental setup for the random Raman laser. QWP - quarter wave plate. PBS - polarized beam splitter. BD - beam dump. BS - beam splitter. EM - energy meter. TS - slightly offset 3x telescope to gently focus the beam onto the sample. $\mathrm{BaSO_4}$ - barium sulfate powder with micron sized particles. CL - collection lens set up to image the sample on to the detector. F - filter. D - either an energy meter, spectrometer, or streak camera depending on the desired measurement.}
\label{fig:Conceptual}
\end{figure}

\begin{figure}[p]
\centering
\includegraphics[width=1.0\textwidth]{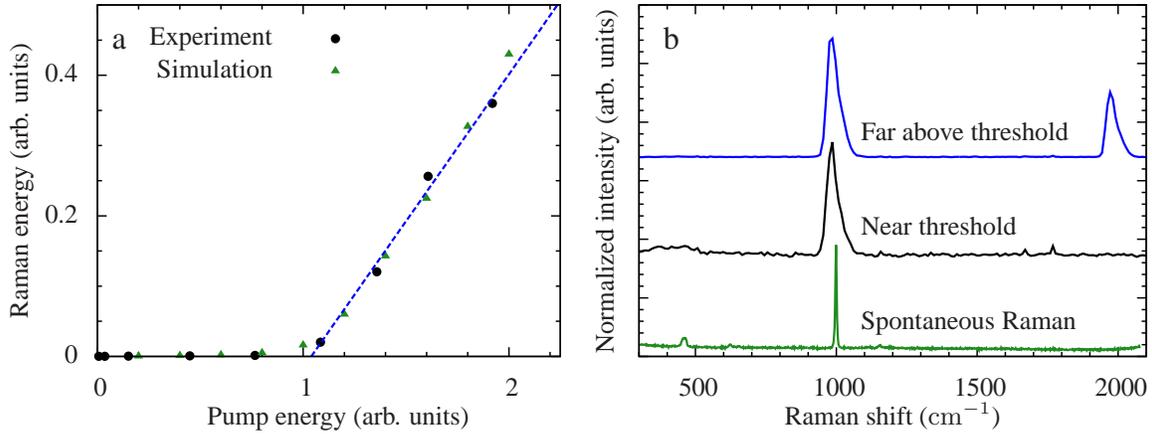}
\caption{\textbf{Random Raman laser efficiency and spectral output.} \textbf{a}, Output Raman pulse energy vs incident pump pulse energy indicating a clear threshold where random Raman lasing begins. Black circles are the experimental data points, dashed blue line is a linear fit of the last 4 experimental data points, and green triangles are the results of Monte Carlo simulations. \textbf{b}, SRS spectra of $\mathrm{BaS0_4}$ powder at two different incident pump powers illustrating the presence of a second order Raman transition, and the spontaneous Raman spectrum of $\mathrm{BaSO_4}$ for comparison. Note, the spontaneous Raman spectrum was adjusted to fit on the same scale as the other two. A higher resolution spectrometer was used for the spontaneous Raman spectrum, which is the reason for the narrower linewidth of this spectrum. The widths of the SRS spectral peaks were limited by the resolution of the spectrometers used.}
\label{fig:Data}
\end{figure}

\begin{figure}[p]
\centering
\includegraphics[width=1.0\textwidth]{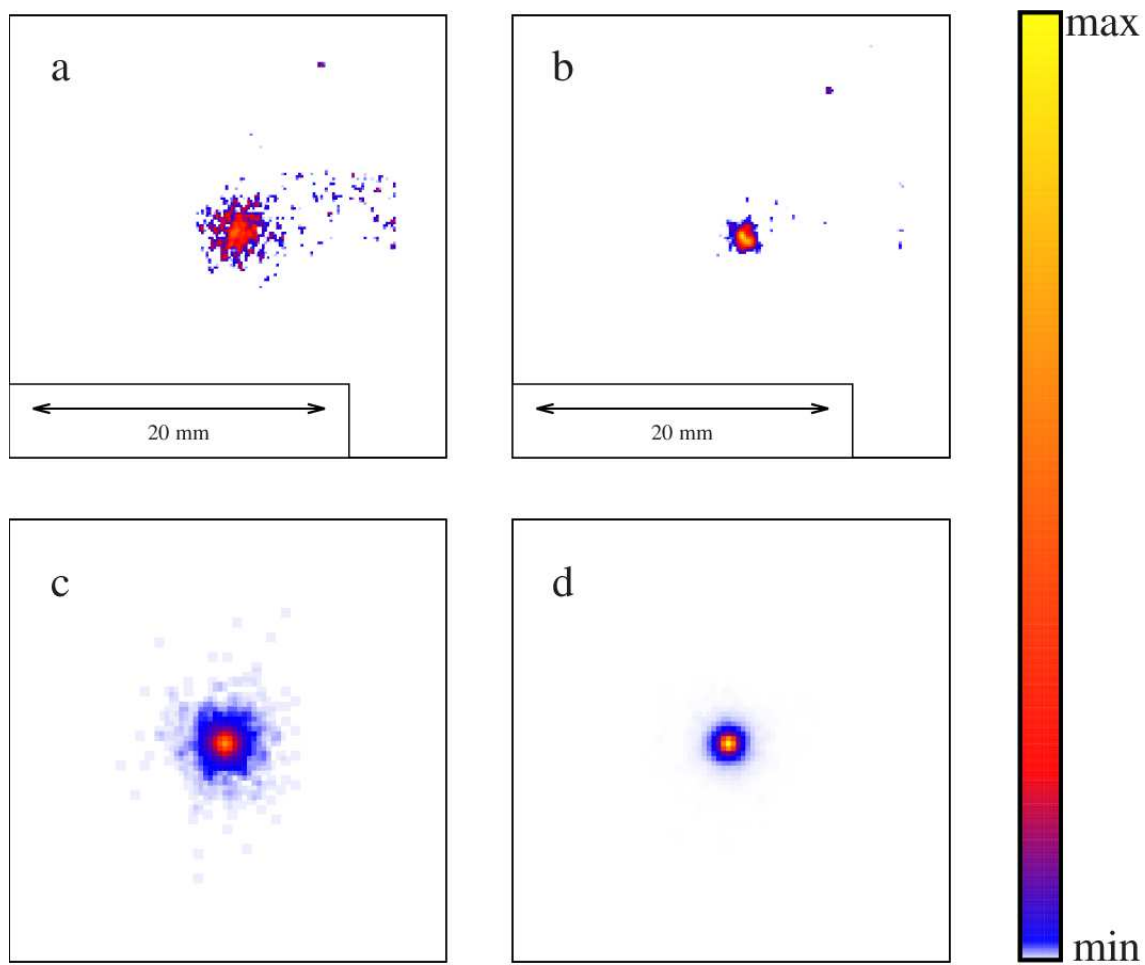}
\caption{\textbf{Spatial profile of the random Raman laser.} \textbf{a}, below threshold experiment, \textbf{b}, above threshold experiment, \textbf{c}, below threshold simulation, and \textbf{d}, above threshold simulation illustrating the significant change in the spatial profile of the Raman signal above and below the threshold. Only background subtraction was preformed for the experimental images.}
\label{fig:Spatial}
\end{figure}
	
\end{document}


\section*{Supplementary Information}

\subsection*{Experimental setup}

For our pump source, we used the second harmonic pulse out of a Spectra Physics Quanta-Ray GCR-3RA that was injection-seeded with a 10 ps pulse out of a Spectra Physics Vanguard HM532. This produces a 40 ps pulse at 532 nm. To control the pump power a quarter wave plate was followed by a polarizing beam splitter. This allows us to adjust the intensity of the pumping laser without effecting the beam quality. The pump pulse is then gently focused onto the 1-5 $\mathrm{\mu m}$ $\mathrm{BaSO_4}$ powder that was loosely packed into a small Petri dish by using a slightly offset 3x telescope consisting of a $150 \; \mathrm{mm}$ focal length plano-convex was followed by a $-50 \; \mathrm{mm}$ focal length plano-concave lens. This setup allows us to slightly tune the spot size on the sample while maintaining a nearly collimated beam. The Fresnel reflections off of a $1.6 \; \mathrm{mm}$ thick BK7 window, tilted at $45^{\circ}$ from the beam axis, was used as a reference signal to measure the pump power. This signal was detected via a Coherent J4-09 energy meter. The reference signal was calibrated by measuring the ratio between the energy of the reference pulse and the energy of the pump pulse at the sample, using an additional Coherent J4-09 energy meter.

The output Raman signal was picked off using a nearly parallel elliptical mirror, and imaged onto the detecting apparatus using a $5.08 \; \mathrm{cm}$ diameter, $125 \; \mathrm{mm}$ focal length lens. The excess $532 \; \mathrm{nm}$ light was removed by using both lenses of a pair UVEX L99-LS6 YAG/KTP laser safety glasses as filters. The two lenses combined to give $9.00 \; \mathrm{OD}$ worth of absorption at $532 \; \mathrm{nm}$, $0.51 \; \mathrm{OD}$ at $563 \; \mathrm{nm}$, and $0.47 \; \mathrm{OD}$ at $594 \; \mathrm{nm}$. These absorption measurements were made using a Cary 5G spectrophotometer. These wavelengths correspond to the pump, the first-order SRS peak at $985 \; \mathrm{cm}^{-1}$, and the second-order SRS peak at $1970 \; \mathrm{cm}^{-1}$ respectively. Using the Ocean Optics USB 2000 spectrometer we confirmed that these filters reduce the $532 \; \mathrm{nm}$ light to a level below that which can be measured by that spectrometer. For measuring the Raman energy in Fig.~2a a Coherent J3S-05 energy meter was used for the detecting apparatus. To measure the SRS spectra shown in Fig.~2b, an Ocean Optics USB 2000 spectrometer was used. For the spontaneous Raman spectrum, an InSpectrum 300-mm spectrometer (Acton Research, Inc.) was used.

The spatial profiles shown in Fig. 3a and 3b were obtained using the same laser system, but the setup differed slightly. Instead of having the powder in a Petri dish and shining the laser down from the top, the $\mathrm{BaSO_4}$ powder was placed in a $2 \; \mathrm{cm}$ by $1 \; \mathrm{cm}$ cuvette and imaged from the front using a Pulnix TM-6701AN monochrome CCD camera.

\subsection*{Monte Carlo simulations}
For spontaneous Raman scattering we made use of the Monte Carlo model developed previously~[27]. We will quickly recap the model here for clarity. A Gaussian pulse, consisting of $10^5$ photons is sent into a scattering medium characterized by scattering ($l_s$), absorption ($l_a$), and Raman ($l_R$) mean free paths. Photons are propagated for a fixed amount of time during each step of the simulation. This is in contrast to many Monte Carlo schemes that propagate each photon by the distance to its next scattering event, thus causing all photons in the simulation to effectively have different clocks. During each step there is a probability of an event given by
\begin{equation}
P_i = 1 - e^{\Delta r / l_i},
\end{equation}
where i represents elastic scattering, absorption, or Raman scattering, and $\Delta r = (c / n) \Delta t$. Elastic scattering is assumed to be anisotropic and is described by the Henyey-Greenstein probability distribution, characterized by the anisotropy factor $g = \langle \cos{\theta} \rangle$. Spontaneous Raman scattering is assumed to be anisotropic.

To include SRS effects we define an interaction between a pump and Raman photons. For each Raman photon inside a sphere of radius $r_{\mathrm{SRS}}$, centered around the pump photon, there is a binomial probability of converting that pump photon into a Raman photon given by
\begin{equation}
p = 1 - e^{-P_{\mathrm{SRS}} \Delta r}.
\end{equation}
Here, $P_{\mathrm{SRS}}$ is a probability per length used to describe stimulated Raman scattering and is related to the Raman gain coefficient, $G$. When a pump photon is converted the new Raman photon takes the direction of the Raman photon whose coin flip generated the conversion.

For all the simulations presented in this paper the following values for the parameters were used: $l_s = 0.005 \; \mathrm{mm}$, $l_a = 10.0 \; \mathrm{mm}$, $l_R = 200 \; \mathrm{mm}$, $r_{\mathrm{SRS}} = 0.025 \; \mathrm{mm}$, $P_{\mathrm{SRS}} = 0.1 \; \mathrm{mm}^{-1}$, $n = 1.6$, $g = 0.6$, $\Delta r = 0.002 \; \mathrm{mm}$. The sample was assumed to be $ 5 \; \mathrm{mm}$ thick and infinite in the transverse directions to simulate an infinitely thick medium (no photons were transmitted through the sample in any run). The incident pulse was assumed to be a $40 \; \mathrm{ps}$ full-width at half-maximum pulse with a full-width at half-maximum beam diameter of $0.1 \; \mathrm{mm}$ To generate the data in Fig. 2a 25 independent runs were averaged, while Figs. 3c and 3d were generated using only a single run. In order to ensure that SRS did not contribute to the below threshold data presented in Fig. 3c stimulated effects were disabled in this simulation.

The video included in the supplemental material was run with the same parameters as above. The black box is drawn for scale and perspective, and is not a physical boundary. It measures 1 cm by 1 cm in the transverse directions, and is 0.5 cm deep (corresponding to the depth of the simulation). Green spheres of radius $r_{\mathrm{SRS}}$ are drawn to indicate photons at the pump wavelength. Likewise, red spheres are drawn to indicate the Raman scattered photons. The videos on the left and right are of the same simulation, but the video on the left draws both the pump and Raman photons, while the video on the right draws only the Raman photons.